\documentclass[10pt,journal,compsoc]{IEEEtran}

\usepackage{epsfig}
\usepackage{graphicx}
\usepackage{amsmath}
\usepackage{rotating}
\usepackage{authblk}
\usepackage{listings} 
\usepackage[utf8]{inputenc}
\usepackage[english]{babel}
\usepackage{verbatim}

\newcommand{\be}{\begin{eqnarray}}
\newcommand{\ee}{\end{eqnarray}}
\newcommand{\ba}{\begin{array}}
\newcommand{\ea}{\end{array}}
\newcommand{\bmn}{\begin{minipage}}
\newcommand{\emn}{\end{minipage}}
\newcommand{\bv}{\begin{verbatim}}
\newcommand{\ev}{\end{verbatim}}
\newcommand{\bt}{\begin{tabular}}
\newcommand{\et}{\end{tabular}}
\newcommand{\bi}{\begin{itemize}}
\newcommand{\ei}{\end{itemize}}
\newcommand{\btab}{\begin{table}}
\newcommand{\etab}{\end{table}}
\newcommand{\btabs}{\begin{table*}}
\newcommand{\etabs}{\end{table*}}
\newcommand{\btabsw}{\begin{sidewaystable}}
\newcommand{\etabsw}{\end{sidewaystable}}
\newcommand{\bfig}{\begin{figure}}
\newcommand{\efig}{\end{figure}}
\newcommand{\bfigs}{\begin{figure*}}
\newcommand{\efigs}{\end{figure*}}
\newcommand{\bc}{\begin{center}}
\newcommand{\ec}{\end{center}}
\newcommand{\bit}{\begin{itemize}}
\newcommand{\eit}{\end{itemize}}

\newcommand{\Sum}{\displaystyle\sum\limits}

\begin{document}

\title{A comprehensive statistical study of metabolic and protein-protein interaction network properties}

\author[1]{D.~Gamermann\thanks{danielg@if.ufrgs.br}}
\author[2]{J.~Triana}
\author[3]{R.~Jaime}

\affil[1]{Department of Physics, Universidade Federal do Rio Grande do Sul (UFRGS) - Instituto de Física \\ Av. Bento Gonçalves 9500 - Caixa Postal 15051 - CEP 91501-970 - Porto Alegre, RS, Brasil.}
\affil[2]{Universidad Politécnica Salesiana - sede de Guayaquil, Ecuador}
\affil[3]{Universidade Hermanos Saíz Montes de Oca - Pinar Del Rio, Cuba}

\maketitle

\begin{abstract}
Understanding the mathematical properties of graphs underling biological systems could give hints on the evolutionary mechanisms behind these structures. In this article we perform a complete statistical analysis over thousands of graphs representing metabolic and protein-protein interaction (PPI) networks. First, we investigate the quality of fits obtained for the nodes degree distributions to power-law functions. This analysis suggests that a power-law distribution poorly describes the data except for the far right tail in the case of PPI networks. Next we obtain descriptive statistics for the main graph parameters and try to identify the properties that deviate from the expected values had the networks been built by randomly linking nodes with the same degree distribution. This survey identifies the properties of biological networks which are not solely the result of their degree distribution, but emerge from yet unidentified mechanisms other than those that drive these distributions. The findings suggest that, while PPI networks have properties that differ from their expected values in their randomized versions with great statistical significance, the differences for metabolic networks have a smaller statistical significance, though it is possible to identify some drift.
\end{abstract}

\begin{keyword}
graphs, biological networks, degree distribution, PPI networks, metabolic networks, scale-free networks
\end{keyword}


\section{Introduction}

Networks are an intuitive way to pictorially represent elements and their interactions in many complex systems. On top of its visual appeal, graph theory is a well established mathematical field which allows these structures to be quantitatively analyzed and have their properties objectively evaluated. As soon as the abstract graph theory began to be applied in order to describe real world networks, it became clear that graphs representing real systems differed vastly from what would be expected from the naive Ërdos-Rénny random model for networks studied in the fifties \cite{erdos}.

Graphs representing many different real world systems such as for example, author citations relations \cite{citenet}, biological networks \cite{protnet1, bionet2} or flight connections \cite{flightnet} present many similar topological properties that significantly distinguish them from Ërdos-Rénny random graphs. Some of these common characteristics, often claimed to be ubiquitous in real world networks, are the presence of hubs (a few highly connected nodes), the so called small world property \cite{Amaral}, scale-freeness or self-similarity as a consequence of the network's nodes degree distribution often be similar to a power-law function ($p(k)\propto k^{-\gamma}$) \cite{barabasi1, barabasi2, rekaalbert}, high clusterization and hierarchical organization \cite{Ravasz}. 

Biological systems are the result of evolution and natural selection. Therefore, the characteristics one observes in biological networks are indications of the evolutionary pressures under which these systems developed. The dense tails of the degree distributions in these networks, for example, have been suggested to give rise to the robustness of these systems against random node deletions \cite{jeong, robust, robust2}, which would reflect the fact that live organisms are resilient to random mutations or to deprivation situations. In fact, many works study evolutionary models that attempt to generate graphs with similar degree distributions than the ones observed in real systems. These models define simple rules for the growth of a network and study the assymptotic behaviour of an evolving graph. In the preferential attachment model \cite{barabaa, barabasi3}, for example, a new node is more likely to connect to already highly connected nodes in the network, while in the duplication and divergence model \cite{Vazquez_2003} an existing node is first copied with the same connections and then may have its connections altered. These works usually focus in obtaining a set of rules that will generate random graphs with degree distributions similar to the ones observed in real systems. Therefore, it is important to determine the characteristics of these distributions on one hand and whether the degree distribution alone is enough to replicate other common characteristics of these networks such as high clusterization.

Though it is a very common claim that biological networks are scale-free (meaning that their node's degree distribution follows a power-law function), there are some studies that dispute this conclusion \cite{khanin, gipsi}. Most works that fit a power-law to the degree distribution of a given network, overlook the quality of the fit. However, in \cite{powerlaw}, an objective statistical study of several different real world networks is made in order tho tackle this issue. In the study, the authors use robust statistical tools not only to fit the distributions but also to evaluate the quality of the fits (p-value), shedding light on which networks representing real systems might and might not be called scale-free. In order to do that, the authors in \cite{powerlaw} chose, for each real world system, a single representative network and analyze the fit obtained for its degree distribution. Though a single metabolic and a single protein-protein interaction (PPI) network are analyzed in the paper, these data sets composed of single elements are not enough to extract strong general conclusions about these biological systems. In \cite{vitor} a version of the duplication and divergence model is analytically studied and it is observed that although it is possible to find parameter regions where the degree distribution will have a denser right tail, this distribution has a peak for low degrees indicating that it can not be fitted to a monotonically decreasing power-law function if one considers its whole range.

The present study has performed two systematical analysis over two huge data sets of biological networks: more than 3000 metabolic and over 1000 PPI networks. The objective of the first analysis is to determine whether the degree distributions of the networks as a whole or their right tails may be well described as scale-free (power-law) functions. For this, we use the same tools as in \cite{powerlaw} to fit the degree distribution of each network to a power-law and standard statistical tools in order to assess the quality of the obtained fits. In a second analysis, the main graph attributes of each network are evaluated, which allows us to draw a general picture about the characteristics of these graphs for a wide range of organisms. In order to identify which properties are simply expected given the degree distributions of the networks and which ones are the result of other\footnote{Not only driving the degree distribution.} possible evolutionary pressures over the underlying biological systems, the same properties are computed over randomized versions of each network that preserve their degree distributions. With this study, it is possible to evaluate the differences between the property in the real networks with respect to their randomized versions, to assess the statistical significance on the existence of such differences and, therefore, to identify attributes that are not only a consequence of the network's degree distribution. 

The work is organized as follows: in the next section we define the graphs we study and from where the data in order to build them was retrieved. The section after that is dedicated to describe all the analysis done. Finally, we present the results and discussion and in the last section a brief overview and our conclusions. We also include an appendix on the distribution of a network characteristic in its randomized versions.


\section{Data and Graphs}

In this section we define the graphs we analyze and describe the procedures followed in order to obtain the data from which we build the networks.

\subsection{Metabolic Networks}

For an organism, we define its metabolic network as the undirected and unweighted graph resulting from connecting the molecules or metabolites appearing in its metabolism based on the biochemical reactions that keep its cells (or cell) alive. Two metabolites (nodes in the graph) are connected if they appear as a substrate-product pair in any chemical reaction in its metabolism.

Therefore, the data needed to build the metabolic network for one organism is the list of all biochemical reactions that can be found in its metabolism. This data was obtained from the Kyoto Encyclopedia of Genes and Genomes (KEGG) database \cite{kegg, kegg2}. First, a list of all genes in an organism is obtained, from this list of genes, those annotated as coding for enzymes are identified, as well as the corresponding chemical reactions catalyzed by each enzyme. In a complementary step, the pathways identified in each organism are obtained and the corresponding \texttt{KGML} files (KEGG Markup Language) are retrieved. These files allow one to identify the non-enzymatic reactions associated with known pathways. After the retrieval process, one has a list of chemical reactions from which one builds the network by listing all single metabolites appearing in the reactions and stipulating an undirected link between two metabolites whenever they appear in opposite sides of a reaction (as substrate-product).

An automated python script which connects to the database rest API was written, in order to obtain KEGG's list of organisms and run through it retrieving the data needed to build the networks \cite{raymari}. The networks for 3481 organisms were successfully build by the procedure.

\subsection{PPI networks}

In a protein-protein interaction network, every protein found in an organism's proteome represents a graph node and two nodes are linked if the proteins have some kind of interaction between them. 

Data for the production of PPI networks was downloaded from the STRING database \cite{string}. From this database, for hundreds of organisms, one obtains lists of pairs of proteins present in the organisms and several scorings for each pair representing the confidence on the existence of some interaction between them (different scores are associated to different sources of evidence for the existence of the interaction). For all organisms downloaded, we built the network for each organism by setting the threshold on the minimal confidence level that an interaction must have in order to define an undirected link in the network. The threshold considered was 0.90 (in a range between 0 and 1) for the combined score.

With this procedure we built 1073 PPI networks.


\section{Statistical Analysis} \label{statana}

In this section we explain the graph parameters and characteristics that were analyzed in each network and the statistical tools employed in the analysis. 

The theory on measurements related to graphs and the study of network characteristics and parameters can be found in several books and reviews. See, for example, \cite{barabasi3, estrada}.

\subsection{Degree Distribution}

It is often claimed that biological networks have scale-free (power-law) node degree distribution. The discrete power-law distribution has the form:

\be
p(x/ \gamma, x_0) &=& \left\{ \ba{cl} \frac{x^{-\gamma}}{\zeta(\gamma, x_0)}  & x\ge x_0 \\
                         0                                        & x< x_0 \ea \right. \\
\zeta(\gamma, x_0) &=& \sum_{x=x_0}^\infty x^{-\gamma}
\ee
where $\zeta(\gamma, x_0)$ is the Riemann zeta function (modified such that the sum starts at a minimum value $x_0$). This distribution has $\gamma$ and $x_0$ as parameters. The parameter $x_0$ is an integer indicating the smallest number in the distribution. 

According to the above, we attempt to fit a power-law (scale-free) distribution to the nodes degree distribution of the studied networks. In order to do that, given the set of $N$ numbers $\{k_i, i= \textrm{1, 2, ..., }N\}$ representing the degree of every node in a network, we find the value of the parameter $\gamma$ that maximize the likelihood\footnote{Since the logarithm is a monotonically increasing function, its maximum is at the same point as the maximum value of its argument.}, for a given value of $x_0$:

\be
\ln {\cal L} &=& \sum_{i=1}^{N^\prime}\ln p(k_i/ \gamma, x_0)
\ee
where the sum is made over the degrees of every node in the network bigger than $x_0$. To find the parameter $\gamma$ that maximizes the likelihood, one must solve the equation:

\be
\frac{d}{d\gamma} \ln {\cal L} &=& 0 \\
N^\prime\frac{\frac{d}{d\gamma}\zeta(\gamma, x_0)}{\zeta(\gamma, x_0)} - \sum_{i=1}^{N^\prime} \ln k_i &=& 0 \label{maxlkhdeq}
\ee
where $N^\prime$ is the number of nodes in the network with degree bigger or equal to $x_0$.

Once $\gamma$ is established, for many possible values $x_0$, we evaluate the goodness of fit through a $\chi^2$ test: the $\chi^2$ statistic is calculated and the right-cumulative distribution of the Pearson's $\chi^2$ distribution at this point is obtained. The result is the p-value i.e. the probability of obtaining a statistical fluctuation bigger than the observed one if the $k_i$'s distribution does come from a power-law with parameters $\gamma$ and $x_0$. Therefore, big values of the p-value indicate a good fit.

For each network, we follow the same procedure: having its degree distribution, for every possible value of $x_0$ between 1 and some\footnote{As will be clear in the Results section, the p-values increase until some given value of $x_0$ and decrease afterwards.} $x_{max}$ we solve Eq. (\ref{maxlkhdeq}) and find the value of $\gamma$ that maximize the likelihood for the given $x_0$. We also evaluate the p-value and count the amount (fraction) of nodes in the network with degree smaller than $x_0$ (these nodes did not participate in the fit procedure). We also evaluate an upper and lower uncertainty for the $\gamma$ parameter by finding the two points around the maximum likelihood where it decreases half point (0.5) \cite{mxlkhdunc}.

\subsection{Graph Properties} \label{graphprop}

For every single graph produced (3481 metabolic networks and 1073 PPI networks), first the values for basic network parameters are obtained. Most of the obtained graphs contained small disconnected components, so we also count, for each graph, the number of disconnected components, the size of the biggest component and the average size of the smaller components. The most straight forward properties that are obtained from the graphs are its number of nodes $N$ and its number of links ${\cal N}$. 

Also, for every network it is straightforward to obtain its node's degree distribution i.e. $n_i$, the number of nodes in each network that have $i$ links, for every possible integer $i$. This is evaluated by first obtaining the degree for each node $i$, $k_i$ which is the number of links node $i$ has:

\be
k_i &=& \sum_{j=1}^N M_{ij}
\ee
where $M_{ij}$ is the adjacency matrix of the graph (a square symmetrical $N\times N$ matrix where each element $M_{ij}$ is 1 if node $i$ is connected to node $j$ and 0 otherwise).

Another local property of each node is its clustering coefficient $C_i$:

\be
C_i &=& \frac{2 E_i}{k_i(k_i-1)}
\ee
where $E_i$ is the number of links between the neighbors of node $i$. This coefficient is the ratio between the number of triangles node $i$ actually forms with its neighbors and the total number of all possible ones given its degree $k_i$. 

The local parameters (properties associated with each node in a network) can be averaged over all nodes in the network in order to establish an average network parameter. In the case of the two above mentioned parameters, one has the network's average degree $\bar{k}$ and average local clustering coefficient $\bar{C}$. It is also possible to define a global parameter representing the clustering of a network as the ratio between all triangles (size 3 clicks\footnote{A click is a complete subgraph of the network.}) the network (as a whole) actually has and the number of possible triangles it could have, based on the number of connected triples (length 2 paths):

\be
{\cal C} &=& 3\frac{|C_3|}{ |P_2|}
\ee
where $|C_3|$ is the number of triangles (tree nodes connected in a cycle) and $|P_2|$ the number of 2-paths (connected triples).

We also study two parameters related to node correlations, namely nodes distances and network assortativity. 

First we evaluate the symmetric distance matrix, a matrix where every element $d_{ij}$ is shortest path length between node $i$ to node $j$, via the Dijkstra's algorithm \cite{dijkstra}. Since we consider the unweighted network (every link has weight 1), the size of the path is set as the distance between the two nodes. The average of all elements\footnote{Since some networks have disconnected components, the distance between nodes in different components is infinity (one is not reachable from the other). These distances are left out from the sum evaluating the average.} in the distance matrix is the network's average distance $\bar{d}$:

\be
\bar{d} &=& \frac{2}{N(N-1)}\Sum_{i=1}^N\Sum_{j>i} d_{ij}
\ee

The network's assortativity, $A$, is a correlation coefficient between the node's excess degree and its expected value in an Ërdos-Rénny random network:

\be
A &=& \frac{1}{\sigma^2_q}\Sum_{k_i, k_j}k_i k_j\left(e(k_i, k_j)-q(k_i)q(k_j)\right) \\
q(k_i) &=& \frac{(k_i+1)p(k_i+1)}{\bar{k}}
\ee
where $p(k_i)=\frac{n_{k_i}}{N}$ is the probability that a node has degree $k_i$, $e(k_i, k_j)$ is the fraction of links in the network connecting nodes of degree $k_i$ with $k_j$, $q(k_i)$ is called the excess degree distribution and $\sigma_q$ is its standard deviation.

Positive coefficient $A$ means an assortative network i.e. high degree nodes tend to be connected to other high degree nodes; while a negative coefficient $A$ means a dissortative network that is, a network where high degree nodes tend to connect with low degree nodes.

\subsection{Network Randomization} \label{randana}

We want to distinguish properties of the network that come solely as a natural consequence of its degree distribution and those that require some underlying mechanism to be achieved. To accomplish that, after having evaluated the network parameters, we compare those to averages of the same parameters evaluated over sets of randomized networks i.e. networks with the same degree distribution for their nodes, but where the links have been randomly exchanged. In the appendix we discuss the distribution of the network properties over the population of networks generated by this randomization process.

The randomization process we implemented is the following: first two links of the network are randomly selected. The links are broken and the nodes participating in one of the original links are connected to the nodes participating in the other original link. Since we work with simple undirected networks, sometimes this process fails, because given the randomly selected links, the relinking of the network would either generate a node connected with itself or two nodes sharing multiple connections. Repeating multiple times this process, one can also estimate (by bootstrap) the amount of times the process failed estimating, in this way, the probability of success in the process, which will be a property solely of the degree distribution of the network. We call this parameter $\xi$. Note that this randomization process does not change the degree distribution of the network i.e. every node keeps its $k_i$ constant during the process.

One is left to decide how many times to repeat this randomization process in order to obtain a truly random network\footnote{Note that this randomization process does not generate a Ërdos-Rénny random network, but a network with the exact same degree distribution as the real original graph.}. We adopt the paradigm that each link should have a 99.9\% probability of having been touched at least once by the process. The idea behind this process is to generate random networks with the same degree distribution as the original network therefore, assessing the properties of these random networks, one obtains the characteristics of the network that emerge only as a consequence of the degree distribution of the graphs. This set of random networks mimics what would be expected as result from an evolutionary model constructed in order to generate networks adjusting the degree distribution observed in real world graphs and not caring with any other aspect of the resulting graphs.

Since in each step of the randomization process two links are selected, the probability $p$ that any given link is touched in a given step is $p=\frac{2}{\cal N}$. Therefore, the probability that a link is not touched is $\bar{p}=1-p$. If the randomization process is repeated $n$ times, the probability that a given link is never touched is $\bar{p}^n$. So the number of times we must repeat the randomization process in order that there is a 99.9\% probability that any given link was touched by the process (probability $\frac{1}{1000}$ of not being touched) is:

\be
n &=& -\frac{\ln(1000)}{\ln\left(1-\frac{2}{\cal N}\right)}
\ee
to give an idea of his number, in a typical metabolic network with 2400 links, this value is $n=8286$, while for a typical PPI network with 28000 links, this number is $n=96705$.

For each network, we obtain 10 randomized versions\footnote{The calculations become computationally intensive for big networks (some PPI networks have over 10000 nodes) and therefore, choosing a bigger number of random samples, would result in unreasonable running time for the calculations over the entire data set.} of it, evaluate each network parameter in each randomized network and estimate an expected value and its uncertainty by evaluating the average and standard deviation of the parameter over the ten randomized network samples. In this way, we are able to evaluate, for each network, a measure of the parameter deviation in the real network from the expected value in the randomized versions of it, by computing the statistical $t_P$ for each parameter $P$:

\be
t_P = \frac{\bar{P}_{random}-P_{real}}{\frac{S}{\sqrt{10}}}
\ee
where $P_{real}$ is the value obtained for the parameter $P$ in the real network, $\bar{P}_{random}$ is the average value of the parameter over the ten randomized versions of the network and $S$ the standard deviation of the parameter in the random samples. 

The statistical $t_P$ can be used to assess the statistical significance on the existence of a difference between the parameter value in the real network and its expected value in randomized versions of it, by evaluating the cumulative student's t-distribution with 9 degrees of freedom at point $|t_P|$. In the appendix we provide a normality test for the distribution of the network characteristics over its randomized versions, justifying thus the use of the student's t-test. Two times the value of this cumulative distribution is interpreted as the p-value for the null hypothesis that the observed network has the $P$ parameter equal to its expected value. High values of this p-value would indicate that the difference is not significant (high probability of obtaining a fluctuation equal or bigger than the observed one in the population). Parameters for which the p-value is small would indicate a significant difference between the observed and expected value, hinting that evolution favors (selects) networks in which the parameter value is bigger (if $t_P$ is negative) or lower (if $t_P$ is positive) than what is expected in a random version of the network. Therefore, a good dynamical evolutionary model for these biological systems would have to incorporate mechanisms that result in networks with such characteristics. 

%


\section{Results and Discussion}

For each data set of networks studied (metabolic and PPI), we present first the results for fitting procedure for the node's degree distributions followed by the descriptive statistics for the network parameters and then the results from the comparison between the real and the randomized network versions.

\subsection{Metabolic Networks}

In table \ref{tab3} we show the results for the fitting procedure done over the node's degree distributions, averaged over the 3481 metabolic networks. For each network, for each value of $x_0$ between 1 and 15, we evaluate the value for the parameter $\gamma$ that maximizes the likelihood for the observed degrees in the network (solution of eq. (\ref{maxlkhdeq})), we obtain its uncertainties, and for the fitted value of $\gamma$ the statistical $\chi^2$ is computed along with its correspondent p-value. Note that the fit procedure is done for every $x_0$ in the table for each one of the 3481 graphs in the data set. Therefore, the values of $\gamma$ and its uncertainties presented in the table are the result of averaging the obtained values of $\gamma$ and its uncertainty in each network weightening the values by the p-value of the fits (giving more importance to the better fitted values). Next to the average p-value we also show the fraction of the fits for which the p-value was bigger than 1\%. Note also that, for each value of $x_0$ bigger than 1, some nodes (the ones for which $k_i<x_0$) are left out of the fit. The table presents also the average fraction of discarded nodes in the fits, for each value of $x_0$.

\btab[h]
\caption{Fitted parameter and uncertainties averaged using p-value as weight, for metabolic networks. In parenthesis, next to the average p-value, the fraction of fits for which the individual p-value was greater than 0.01.} \label{tab3}
\bc
\scalebox{0.8}{\bt{c|ccc}
$x_0$ & $\gamma$ & Discarded & average p-value  (fraction $>$ 0.01)\\
\hline
 & & \\
1 & $1.569968^{+0.057497}_{-0.053732}$ & 0.000000 & $0.000000 \pm 0.000000$ (0.000000 \%) \\
 & & \\
2 & $2.134354^{+0.061622}_{-0.059194}$ & 0.045531 & $0.012007 \pm 0.037116$ (58.527493 \%) \\
 & & \\
3 & $2.372021^{+0.139983}_{-0.130618}$ & 0.358368 & $0.003266 \pm 0.026371$ (12.395154 \%) \\
 & & \\
4 & $2.654597^{+0.214409}_{-0.196739}$ & 0.519733 & $0.001849 \pm 0.021633$ (6.337372 \%) \\
 & & \\
5 & $2.592453^{+0.251536}_{-0.226789}$ & 0.693625 & $0.001751 \pm 0.018163$ (7.362535 \%) \\
 & & \\
6 & $2.730658^{+0.324567}_{-0.287168}$ & 0.761719 & $0.001357 \pm 0.015890$ (6.150979 \%) \\
 & & \\
7 & $2.700261^{+0.356084}_{-0.311195}$ & 0.820451 & $0.001273 \pm 0.013544$ (7.362535 \%) \\
 & & \\
8 & $2.797779^{+0.407363}_{-0.352410}$ & 0.849825 & $0.000934 \pm 0.010066$ (5.498602 \%) \\
 & & \\
9 & $2.794249^{+0.449053}_{-0.383302}$ & 0.880596 & $0.000782 \pm 0.007931$ (4.473439 \%) \\
 & & \\
10 & $2.863253^{+0.504359}_{-0.425526}$ & 0.898041 & $0.000683 \pm 0.006580$ (4.287046 \%) \\
 & & \\
11 & $2.910327^{+0.552005}_{-0.460801}$ & 0.911973 & $0.000647 \pm 0.005911$ (4.380242 \%) \\
 & & \\
12 & $3.006127^{+0.618104}_{-0.509971}$ & 0.921416 & $0.000545 \pm 0.005157$ (3.914259 \%) \\
 & & \\
13 & $3.102794^{+0.690538}_{-0.562881}$ & 0.929619 & $0.000455 \pm 0.004476$ (3.075489 \%) \\
 & & \\
14 & $3.166999^{+0.753881}_{-0.607058}$ & 0.936420 & $0.000414 \pm 0.004488$ (2.423113 \%) \\
 & & \\
15 & $3.429787^{+0.859267}_{-0.689126}$ & 0.937253 & $0.000431 \pm 0.005674$ (2.329916 \%) \\
 & & &
\et}
\ec
\etab

For $x_0=1$, the p-value is lower than $10^{-6}$, indicating that only the tail of the distribution might be well adjusted to a power-law function and the highest p-values obtained are for $x_0=2$. In this case, the value of $\gamma$ is a little above 2.1 and around 4\% of the nodes in the network have degree 1 and do not participate in the fit. But even in this case, the average p-value is around 0.01 indicating that the deviation from a power-law is important and there is small probability of observing such fluctuation if the degrees came from the hypothesized (scale-free) distribution.

Now, for each one of the 3481 metabolic networks in our data set, we evaluated the graph properties and characteristics described in subsection \ref{graphprop}. Histograms depicting the distributions of the main parameters over our data set are shown in figure \ref{fig1}. The distributions show the bulk of the data distributed around a central value, but all of them also present a significant number of outliers.

\bfigs[h]
\bc
\bt{cc}
\scalebox{0.5}{\input{plots/N}} & \scalebox{0.5}{\input{plots/calN}} \\
\scalebox{0.5}{\input{plots/bark}} & \scalebox{0.5}{\input{plots/barC}} \\
\scalebox{0.5}{\input{plots/bard}} & \scalebox{0.5}{\input{plots/A}} 
\et
\ec
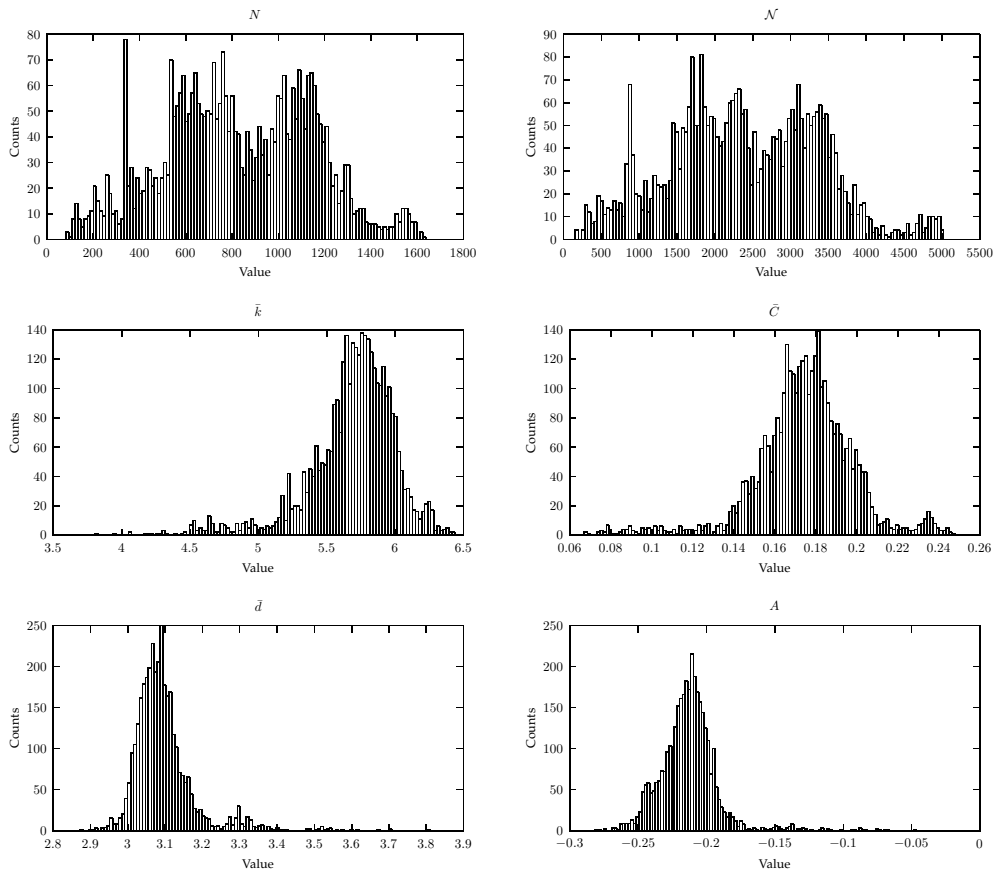
\caption{Histograms for the parameter distributions in metabolic networks. Top left, right: number of nodes and number of links. Center left, right: average degree and average local clustering. Bottom left, right: Average shortest path and assortativity.} \label{fig1}
\efigs

In table \ref{tab1} we present the descriptive statistic for the parameters. Since some distributions have a sizable skewness (asymmetry), besides evaluating the standard deviation of the distribution, we also evaluated the standard deviation for all values bigger and smaller than the average, separately. These are shown in the table as uncertainties around the average value of each parameter. 

\btabs[h]
\caption{Descriptive Statistics for the distribution of metabolic network parameters. The first two columns show the parameter name (definition) and its symbol as used in the present work. Around the average, shown as uncertainties, are the standard deviations calculated for values bigger and smaller than the average, separately.} \label{tab1}
\bc
\bt{c|c|c|c|c}
Parameter & Math Symbol & Average & Standard Deviation & Skewness \\
\hline
     &         &        &           &     \\
  Number of nodes  &   $N$   &   $827.228670^{+327.515437}_{-313.031642}$   &   319.999188  &  0.000736   \\
     &         &        &           &     \\
  Number of links  &   ${\cal N}$   &   $2384.474864^{+1017.416129}_{-951.857140}$   &   983.598923  &  0.093452   \\
     &         &        &           &     \\
  Average degree  &   $\bar{k}$   &   $5.686846^{+0.248550}_{-0.400098}$   &   0.321772  &  -1.359436   \\
     &         &        &           &     \\
  Average local clustring  &   $\bar{C}$   &   $0.174469^{+0.021995}_{-0.026009}$   &   0.024008  &  -0.658240   \\
     &         &        &           &     \\
  Global clustering  &   ${\cal C}$   &   $0.052559^{+0.014486}_{-0.011211}$   &   0.012855  &  0.776866   \\
     &         &        &           &     \\
  Assortativity  &   $A$   &   $-0.213441^{+0.022995}_{-0.018595}$   &   0.020923  &  1.587561   \\
     &         &        &           &     \\
  Average distance  &   $\bar{d}$   &   $3.094789^{+0.110611}_{-0.054124}$   &   0.081075  &  2.340936   \\
     &         &        &           &     \\
  Average number  &   $P_2$   &   $104553.170928^{+91543.797126}_{-59024.598610}$   &   74647.830744  &  1.090137   \\
  of 2-paths   &                   &                &           &     \\
  Average number  &   $C_3$   &   $1564.592933^{+916.729975}_{-748.085279}$   &   830.036302  &  0.520495   \\
  of triangles   &                   &                &           &     \\
  Average number  &   $N_{comps}$   &   $7.202528^{+5.224926}_{-3.526826}$   &   4.316908  &  0.762602   \\
  of components   &                   &                &           &     \\
  Average size of  &   ${\cal S}_{main}$   &   $813.934502^{+318.963107}_{-306.925997}$   &   312.720429  &  -0.008834   \\
  main Component   &                   &                &           &     \\
  Average size  &   ${\cal S}_{small}$   &   $2.053929^{+0.487501}_{-0.470099}$   &   0.475511  &  -2.004091   \\
  smaller components   &                   &                &           &     
\et
\ec
\etabs

In a real metabolic network, one would not expect disconnected components. The small components into which the networks fragment themselves are possibly a problem of wrong annotations in the databases or misidentification of some chemical reactions or metabolites within them in the automated process of reconstructing the networks. In any case, as can be seen from the difference between the average size of the main components and the average total number of nodes, the disconnected components amount to a negligible number of nodes. 

All metabolic networks are dissortative ($A<0$). It is also possible to observe that, while the networks show a high average local clustering, their global clustering parameter tend to be small, close to zero. The networks tend to cluster locally, but not globally. The lack of correlation between the local and global clustering coefficients in real networks has already been observed in other systems \cite{estrada, newman}.

Table \ref{tab1rand} shows the same parameters as in table \ref{tab1} (except for those that are not affected by the randomization process, like $N$, $\cal N$ or $P_2$) evaluated for the average values of the randomized samples of each network. The last row in this table presents the parameter $\xi$, whose average value is around 0.87, indicating that, on average, in 13\% of the randomization steps, the links broken could not have been properly relinked (the random step failed).

\btabs[h]
\caption{Parameter values in randomized versions of the metabolic networks.} \label{tab1rand}
\bc
\bt{c|c|c|c|c}
Parameter & Math Symbol & Average & Standard Deviation & Skewness \\
\hline
     &         &        &           &     \\
  Average local clustring  &   $\bar{C}_{rand}$   &   $0.184840^{+0.021231}_{-0.025024}$   &   0.023123  &  -0.548522   \\
     &         &        &           &     \\
  Global clustering  &   ${\cal C}_{rand}$   &   $0.049392^{+0.014380}_{-0.009870}$   &   0.012134  &  1.112934   \\
     &         &        &           &     \\
  Assortativity  &   $A_{rand}$   &   $-0.206991^{+0.020612}_{-0.015880}$   &   0.018317  &  1.247442   \\
     &         &        &           &     \\
  Average distance  &   $\bar{d}_{rand}$   &   $3.013782^{+0.057361}_{-0.050978}$   &   0.054262  &  0.760095   \\
     &         &        &           &     \\
  Average number  &   $C_{3rand}$   &   $1469.290822^{+875.345398}_{-705.010822}$   &   787.625049  &  0.521137   \\
  of triangles   &                   &                &           &     \\
  Average number  &   $N_{compsrand}$   &   $1.255544^{+0.263182}_{-0.162395}$   &   0.208086  &  0.892551   \\
  of components   &                   &                &           &     \\
  Average size  &   ${\cal S}_{mainrand}$   &   $826.693292^{+327.009958}_{-312.930479}$   &   319.708235  &  0.000727   \\
  main Component   &                   &                &           &     \\
  Average size  &   ${\cal S}_{smallrand}$   &   $0.450593^{+0.380467}_{-0.285910}$   &   0.331312  &  0.586352   \\
  of smaller components   &                   &                &           &     \\
  Probability of success  &   $\xi$   &   $0.866479^{+0.012417}_{-0.011787}$   &   0.012108  &  0.307997   \\
  in the randomization   &                   &                &           &     \\
  process   &                   &                &           &     
\et
\ec
\etabs

The statistical significance of the differences between the parameters in tables \ref{tab1} and \ref{tab1rand} can be appreciated in table \ref{tab1delta} where it is shown the average value for the statistical $t_P$ for each parameter. For each network (3481 in total) the value of $t_P$ and its correspondent p-value is evaluated. The table shows the statics of the distribution of $t_P$ over the whole data set. Next to the average p-value, we also present the fraction of the networks for which this p-value was below 0.05. Histograms for the distribution of $t_P$ for the different parameters can be found in figure \ref{figtmeta}.

\btabs[h]
\caption{Deviation of real metabolic network parameters from the randomized expected values. The column p-value presents the average p-value for the student's t-test evaluated as described in subsection \ref{randana}. In the parenthesis next to it, it is presented the fraction of networks for which the p-value was below 0.05.} \label{tab1delta}
\bc
\bt{c|c|c|c|c}
$t$ & Average & Standard Deviation & Skewness & p-val (fracion $<0.05$) \\
\hline
     &         &        &           &     \\
   $t_{\bar{C}}$   &   $3.998663^{+7.164259}_{-6.293898}$   &   6.723294  &  0.319300  & 0.106481  (74.289 \%) \\
     &         &        &           &     \\
   $t_{\bar{\cal C}}$   &   $-4.565913^{+5.219167}_{-5.353847}$   &   5.285228  &  -0.013612  & 0.086937  (76.300 \%) \\
     &         &        &           &     \\
   $t_{\bar{d}}$   &   $-15.517486^{+6.222337}_{-10.792203}$   &   8.417728  &  -1.588110  & 0.002822  (99.138 \%) \\
     &         &        &           &     \\
   $t_A$   &   $6.882841^{+6.991963}_{-4.492181}$   &   5.685528  &  1.056430  & 0.072785  (81.270 \%) \\
     &         &        &           &     \\
   $t_{C_3}$   &   $-4.565913^{+5.219167}_{-5.353847}$   &   5.285228  &  -0.013612  & 0.086937  (76.300 \%) \\
     &         &        &           &     
\et
\ec
\etabs

\bfigs[h]
\bc
\bt{cc}
\scalebox{0.5}{\input{plots/t_A}} & \scalebox{0.5}{\input{plots/t_bard}} \\
\scalebox{0.5}{\input{plots/t_barC}} & \scalebox{0.5}{\input{plots/t_barcalC}} 
\et
\ec
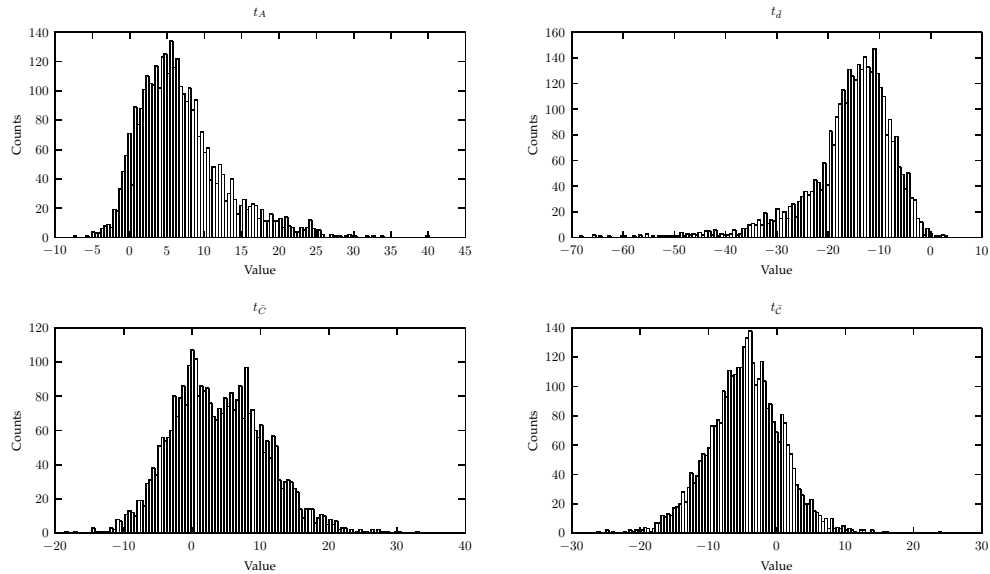
\caption{Histograms for the distribution of the parameter $t_P$ in metabolic networks. Top left, right: assortativity and average shortest path. Bottom left, right: Average local clustering and global clustering.} \label{figtmeta}
\efigs

The most significant difference observed is for the average shortest path. Note, that the significance in this analysis is not on the amount of the difference, but on the confidence level for the existence of this difference. If one compares the difference between $\bar{d}$ and and $\bar{d}_{rand}$ in tables \ref{tab1} and \ref{tab1rand} it amounts to around 2.5\%. Though this difference is small, given any metabolic network, one can say with high confidence that the average shortest path is bigger in the real network than its expected value in randomized versions of it. For the other parameters the situation is not as clear. One observes that a sizable fraction of the networks (around 3/4 of them) do show significance on the existence of some difference (p-val$<0.05$), but these differences are not ubiquitous.

\subsection{PPI Networks}

The same analysis were done over the data set of 1073 protein-protein interaction (PPI) networks. First, in table \ref{tab6} we present the results of the fit procedure for the degree distributions. Here, one can can see that it is not possible to obtain a reasonable fit unless more than 50\% of the network's nodes are left out of the fit. Only for the far right tail of the distribution (around $x_0\geq 13$) one begins to obtain reasonable quality fits. In particular, for $x_0=1$ the fits were so bad, that the p-values were all smaller than the machine precision ($p < 10^{-16}$) and therefore it was not possible to evaluate the averages weighted by the p-values.

\btab[h] 
\caption{Fitted parameter and uncertainties weight averaged by p-value for PPI networks. In parenthesis, next to the average p-value, the fraction of fits for which the individual p-value was greater than 0.01.} \label{tab6}
\bc
\tiny
\bt{c|ccc}
$x_0$ & $\gamma$ & Discarded & average p-value (fraction $>$ 0.01)\\
\hline
 & & \\
2 & $1.865963^{+0.099169}_{-0.091994}$ & 0.192307 & $0.000000 \pm 0.000000$  (0.000000 \%) \\
 & & \\
3 & $2.221444^{+0.151157}_{-0.139448}$ & 0.307692 & $0.000000 \pm 0.000001$  (0.000000 \%) \\
 & & \\
4 & $2.439663^{+0.204404}_{-0.186506}$ & 0.470854 & $0.000000 \pm 0.000003$  (0.000000 \%) \\
 & & \\
5 & $2.832902^{+0.281938}_{-0.255359}$ & 0.545912 & $0.000002 \pm 0.000077$  (0.000000 \%) \\
 & & \\
6 & $3.242113^{+0.375284}_{-0.336876}$ & 0.610980 & $0.000013 \pm 0.000402$  (0.093197 \%) \\
 & & \\
7 & $3.450916^{+0.474926}_{-0.419303}$ & 0.701252 & $0.000015 \pm 0.000455$  (0.093197 \%) \\
 & & \\
8 & $3.898085^{+0.560377}_{-0.491327}$ & 0.682979 & $0.000110 \pm 0.002786$  (0.186393 \%) \\
 & & \\
9 & $3.130937^{+0.321283}_{-0.280358}$ & 0.586882 & $0.000206 \pm 0.003768$  (0.279590 \%) \\
 & & \\
10 & $2.455655^{+0.104328}_{-0.094032}$ & 0.534252 & $0.000896 \pm 0.012494$ (1.025163 \%) \\
 & & \\
11 & $2.439406^{+0.102351}_{-0.089275}$ & 0.559666 & $0.002384 \pm 0.023471$  (1.863933 \%) \\
 & & \\
12 & $2.476708^{+0.090556}_{-0.080428}$ & 0.578328 & $0.005589 \pm 0.045938$  (3.727866 \%) \\
 & & \\
13 & $2.475588^{+0.070015}_{-0.064430}$ & 0.591279 & $0.011535 \pm 0.073171$  (6.150979 \%) \\
 & & \\
14 & $2.505834^{+0.058103}_{-0.056443}$ & 0.599493 & $0.021006 \pm 0.100260$  (11.649581 \%) \\
 & & \\
15 & $2.535557^{+0.059400}_{-0.057562}$ & 0.601718 & $0.037984 \pm 0.132088$  (18.732526 \%) \\
 & & \\
16 & $2.563771^{+0.058223}_{-0.056523}$ & 0.603065 & $0.066005 \pm 0.177740$  (27.772600 \%) \\
 & & \\
17 & $2.607947^{+0.059156}_{-0.056975}$ & 0.612704 & $0.102540 \pm 0.226509$  (32.898416 \%) \\
 & & \\
18 & $2.665189^{+0.062425}_{-0.060167}$ & 0.628512 & $0.141075 \pm 0.269827$  (38.956198 \%) \\
 & & \\
19 & $2.713628^{+0.066267}_{-0.064029}$ & 0.644746 & $0.177258 \pm 0.303147$  (45.013979 \%) \\
 & & \\
20 & $2.765948^{+0.069993}_{-0.066823}$ & 0.660004 & $0.212718 \pm 0.330160$  (49.953402 \%) \\
 & & \\
21 & $2.820009^{+0.074240}_{-0.071932}$ & 0.673400 & $0.248691 \pm 0.351983$  (54.426841 \%) \\
 & & \\
22 & $2.870172^{+0.076718}_{-0.074182}$ & 0.686005 & $0.282034 \pm 0.368536$  (58.434296 \%) \\
 & & \\
23 & $2.921708^{+0.080703}_{-0.078049}$ & 0.697715 & $0.309604 \pm 0.379620$  (62.068966 \%) \\
 & & \\
24 & $2.971699^{+0.085828}_{-0.082439}$ & 0.708591 & $0.338036 \pm 0.388383$  (64.958062 \%) \\
 & & \\
25 & $3.017648^{+0.088027}_{-0.085234}$ & 0.718502 & $0.364349 \pm 0.395045$  (67.567568 \%) \\
 & & \\
26 & $3.063582^{+0.091597}_{-0.088628}$ & 0.727871 & $0.384254 \pm 0.398943$  (69.897484 \%) \\
 & & \\
27 & $3.111638^{+0.095501}_{-0.092096}$ & 0.736455 & $0.401460 \pm 0.401592$  (72.320596 \%) \\
 & & \\
28 & $3.153570^{+0.098928}_{-0.095350}$ & 0.743871 & $0.412428 \pm 0.401466$  (74.650513 \%) \\
 & & \\
29 & $3.193774^{+0.102076}_{-0.097709}$ & 0.750596 & $0.419412 \pm 0.401385$  (75.955266 \%) \\
 & & \\
30 & $3.230413^{+0.105087}_{-0.099465}$ & 0.757299 & $0.424672 \pm 0.401179$  (76.514445 \%) \\
 & & \\
31 & $3.268202^{+0.108429}_{-0.103152}$ & 0.764223 & $0.426257 \pm 0.399990$  (77.446412 \%) \\
 & & \\
32 & $3.302613^{+0.111565}_{-0.104587}$ & 0.770964 & $0.424518 \pm 0.398700$  (78.285182 \%) \\
 & & \\
33 & $3.338181^{+0.115467}_{-0.108414}$ & 0.777333 & $0.419403 \pm 0.398042$  (78.378378 \%) \\
 & & \\
34 & $3.370140^{+0.118337}_{-0.110493}$ & 0.783326 & $0.413641 \pm 0.398115$  (78.657968 \%) \\
 & & \\
35 & $3.399909^{+0.120654}_{-0.113122}$ & 0.788990 & $0.405643 \pm 0.398281$  (78.191985 \%) \\
 & & \\
36 & $3.429541^{+0.123410}_{-0.115413}$ & 0.793904 & $0.395080 \pm 0.397048$  (78.098788 \%) \\
 & & \\
37 & $3.457949^{+0.126062}_{-0.117349}$ & 0.798374 & $0.383044 \pm 0.394640$  (77.912395 \%) \\
 & & \\
38 & $3.483018^{+0.128546}_{-0.117759}$ & 0.802189 & $0.372147 \pm 0.393137$  (76.421249 \%) \\
 & & \\
39 & $3.507460^{+0.131886}_{-0.120585}$ & 0.805173 & $0.358892 \pm 0.390367$  (75.582479 \%) \\
 & & \\
40 & $3.533017^{+0.134452}_{-0.123458}$ & 0.808156 & $0.347501 \pm 0.387838$  (73.904939 \%) \\
 & & & 
\et
\ec
\etab

Next we performed the analysis of the network properties of the data set. In figure \ref{fig2} we present the histograms for the distributions of the main graph parameters and in table \ref{tab4}, the computed descriptive statistics of the distributions. 

\bfigs[h]
\bc
\bt{cc}
\scalebox{0.5}{\input{plots/Np}} & \scalebox{0.5}{\input{plots/calNp}} \\
\scalebox{0.5}{\input{plots/barkp}} & \scalebox{0.5}{\input{plots/barCp}} \\
\scalebox{0.5}{\input{plots/bardp}} & \scalebox{0.5}{\input{plots/Ap}} 
\et
\ec
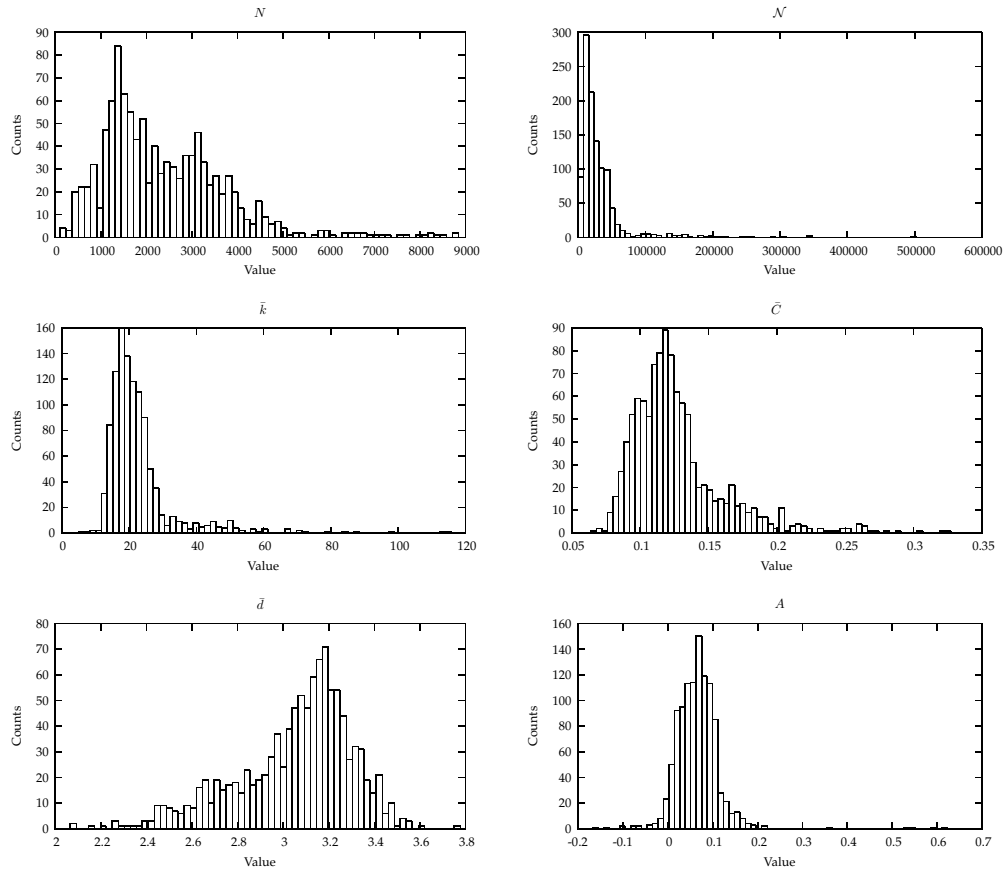
\caption{Histograms for the parameters in PPI networks. Top left, right: number of nodes and number of links. Center left, right: average degree and average clustering. Bottom left, right: Average shortest path and assortativity.} \label{fig2}
\efigs

\btabs[h]
\caption{Descriptive Statistics for the distribution of PPI network parameters.} \label{tab4}
\bc
\bt{c|c|c|c|c}
Parameter & Math Symbol & Average & Standard Deviation & Skewness \\
\hline
  Number of nodes  &   $N$   &   $2378.162162^{+1674.170267}_{-1062.364609}$   &   1360.967079  &  1.312170   \\
     &         &        &           &     \\
  Number of links  &   ${\cal N}$   &   $30648.163094^{+63095.796878}_{-16283.411100}$   &   38129.530762  &  5.244932   \\
     &         &        &           &     \\
  Average degree  &   $\bar{k}$   &   $22.660161^{+16.152767}_{-5.400247}$   &   10.321305  &  3.606450   \\
     &         &        &           &     \\
  Average local clustring  &   $\bar{C}$   &   $0.126916^{+0.048151}_{-0.023306}$   &   0.034739  &  1.852084   \\
     &         &        &           &     \\
  Global clustering  &   ${\cal C}$   &   $0.103259^{+0.077102}_{-0.023378}$   &   0.046880  &  4.257505   \\
     &         &        &           &     \\
  Assortativity  &   $A$   &   $0.065945^{+0.060521}_{-0.042112}$   &   0.051916  &  3.538212   \\
     &         &        &           &     \\
  Average distance  &   $\bar{d}$   &   $3.053442^{+0.200444}_{-0.315897}$   &   0.253305  &  -0.846669   \\
     &         &        &           &     \\
  Average number  &   $P_2$   &   $1998518.199441^{+12364896.830628}_{-1324465.387398}$   &   5674983.876744  &  11.761741   \\
  of 2-paths   &                   &                &           &     \\
  Average number  &   $C_3$   &   $65643.418453^{+378962.677641}_{-44543.188780}$   &   176952.116048  &  11.214598   \\
  of triangles   &                   &                &           &     \\
  Average number  &   $N_{comps}$   &   $3.057782^{+2.966847}_{-1.430859}$   &   2.064812  &  1.279759   \\
  of components   &                   &                &           &     \\
  Average size of  &   ${\cal S}_{main}$   &   $2372.873253^{+1672.813536}_{-1060.363478}$   &   1359.360376  &  1.317361   \\
  Main Component   &                   &                &           &     \\
  Average size  &   ${\cal S}_{small}$   &   $1.915665^{+1.255322}_{-1.919120}$   &   1.455628  &  1.267565   \\
  of smaller components   &                   &                &           &     
\et
\ec
\etabs

The PPI networks show some peculiar properties if compared with the metabolic ones. They present both, local and global clustering coefficients significantly different from zero; These networks are mildly assortative and, though they are bigger in size, they are more densely connected and present a similar shortest average path than the metabolic graphs.

Table \ref{tab2rand} shows the descriptive statistics for the average values of the parameters obtained from the randomized samples. These values show some sizable differences with respect to the parameters in the real networks shown in table \ref{tab4}. For the random samples, the global and local clustering coefficients are close to zero. The different behavior of the global clustering coefficient can be directly linked to the number of triangles in the networks: the real networks present almost twice more size 3 clicks than the random samples. Moreover, the networks present a mild dissortative degree correlation and a slightly smaller average shortest path. The statistical significance of the differences can be read in table \ref{tab2delta}, where we present the descriptive statics for the $t_P$ statisticals and their associated averaged p-values. In figure \ref{fig2rand} are depicted the histograms for the distributions of these statisticals.

\btabs[h]
\caption{Parameter values in randomized versions of the PPI networks.} \label{tab2rand}
\bc
\bt{c|c|c|c|c}
Parameter & Math Symbol & Average & Standard Deviation & Skewness \\
\hline
     &         &        &           &     \\
  Average local clustring  &   $\bar{C}_{rand}$   &   $0.043989^{+0.041802}_{-0.012590}$   &   0.025241  &  3.782684   \\
     &         &        &           &     \\
  Global clustering  &   ${\cal C}_{rand}$   &   $0.043776^{+0.038540}_{-0.012340}$   &   0.023717  &  3.300061   \\
     &         &        &           &     \\
  Assortativity  &   $A_{rand}$   &   $-0.024558^{+0.018546}_{-0.013473}$   &   0.016204  &  2.909341   \\
     &         &        &           &     \\
  Average distance  &   $\bar{d}_{rand}$   &   $2.765340^{+0.201126}_{-0.272347}$   &   0.235211  &  -0.607108   \\
     &         &        &           &     \\
  Average number  &   $C_{3rand}$   &   $28250.808760^{+285093.382085}_{-18602.848604}$   &   106267.234720  &  14.761007   \\
  of triangles   &                   &                &           &     \\
  Average number  &   $N_{compsrand}$   &   $1.352190^{+0.622019}_{-0.255707}$   &   0.425907  &  2.844741   \\
  of components   &                   &                &           &     \\
  Average size of &   ${\cal S}_{mainrand}$   &   $2377.441473^{+1673.836734}_{-1062.029476}$   &   1360.641365  &  1.312565   \\
  Main Component   &                   &                &           &     \\
  Average size  &   ${\cal S}_{smallrand}$   &   $0.529844^{+0.628106}_{-0.387873}$   &   0.507262  &  1.765283   \\
  of smaller components   &                   &                &           &     \\
  Probability of success  &   $\xi$   &   $0.982054^{+0.007652}_{-0.038223}$   &   0.020064  &  -6.207947   \\
  in the randomization   &                   &                &           &     \\
  process   &                   &                &           &     
\et
\ec
\etabs

\btabs[h]
\caption{Deviation of real PPI network parameters from randomized expected values. The column p-value presents the average p-value for the student's t-test evaluated as described in subsection \ref{randana}. In the parenthesis next to it, it is presented the fraction of networks for which the p-value was lower than 0.05.} \label{tab2delta}
\bc
\bt{c|c|c|c|c}
$t$ & Average & Standard Deviation & Skewness & p-val (fracion $<0.05$) \\
\hline
     &         &        &           &     \\
   $t_{\bar{C}}$   &   $-193.127920^{+95.366190}_{-235.414020}$   &   160.654646  &  -3.246060  & 0.000581  (99.814 \%) \\
     &         &        &           &     \\
   $t_{\bar{\cal C}}$   &   $-330.674787^{+181.985206}_{-645.909165}$   &   397.203246  &  -9.187743  & 0.000001  (100.000 \%) \\
     &         &        &           &     \\
   $t_{\bar{d}}$   &   $-190.808070^{+80.353118}_{-173.612148}$   &   123.645141  &  -2.145209  & 0.000098  (99.907 \%) \\
     &         &        &           &     \\
   $t_A$   &   $-47.170522^{+23.227656}_{-57.537909}$   &   39.497893  &  -3.827625  & 0.000728  (99.814 \%) \\
     &         &        &           &     \\
   $t_{C_3}$   &   $-333.068414^{+184.083597}_{-650.350176}$   &   399.742838  &  -9.028683  & 0.000217  (99.907 \%) \\
    &         &        &           &     
\et    
\ec
\etabs

\bfigs[h]
\bc
\bt{cc}
\scalebox{0.5}{\input{plots/t_Ap}} & \scalebox{0.5}{\input{plots/t_bardp}} \\
\scalebox{0.5}{\input{plots/t_barCp}} & \scalebox{0.5}{
\setlength{\unitlength}{0.240900pt}
\ifx\plotpoint\undefined\newsavebox{\plotpoint}\fi
\begin{picture}(1500,900)(0,0)
\sbox{\plotpoint}{\rule[-0.200pt]{0.400pt}{0.400pt}}%
\put(171.0,131.0){\rule[-0.200pt]{4.818pt}{0.400pt}}
\put(151,131){\makebox(0,0)[r]{ 0}}
\put(1419.0,131.0){\rule[-0.200pt]{4.818pt}{0.400pt}}
\put(171.0,223.0){\rule[-0.200pt]{4.818pt}{0.400pt}}
\put(151,223){\makebox(0,0)[r]{ 50}}
\put(1419.0,223.0){\rule[-0.200pt]{4.818pt}{0.400pt}}
\put(171.0,315.0){\rule[-0.200pt]{4.818pt}{0.400pt}}
\put(151,315){\makebox(0,0)[r]{ 100}}
\put(1419.0,315.0){\rule[-0.200pt]{4.818pt}{0.400pt}}
\put(171.0,407.0){\rule[-0.200pt]{4.818pt}{0.400pt}}
\put(151,407){\makebox(0,0)[r]{ 150}}
\put(1419.0,407.0){\rule[-0.200pt]{4.818pt}{0.400pt}}
\put(171.0,500.0){\rule[-0.200pt]{4.818pt}{0.400pt}}
\put(151,500){\makebox(0,0)[r]{ 200}}
\put(1419.0,500.0){\rule[-0.200pt]{4.818pt}{0.400pt}}
\put(171.0,592.0){\rule[-0.200pt]{4.818pt}{0.400pt}}
\put(151,592){\makebox(0,0)[r]{ 250}}
\put(1419.0,592.0){\rule[-0.200pt]{4.818pt}{0.400pt}}
\put(171.0,684.0){\rule[-0.200pt]{4.818pt}{0.400pt}}
\put(151,684){\makebox(0,0)[r]{ 300}}
\put(1419.0,684.0){\rule[-0.200pt]{4.818pt}{0.400pt}}
\put(171.0,776.0){\rule[-0.200pt]{4.818pt}{0.400pt}}
\put(151,776){\makebox(0,0)[r]{ 350}}
\put(1419.0,776.0){\rule[-0.200pt]{4.818pt}{0.400pt}}
\put(171.0,131.0){\rule[-0.200pt]{0.400pt}{4.818pt}}
\put(171,90){\makebox(0,0){-9000}}
\put(171.0,756.0){\rule[-0.200pt]{0.400pt}{4.818pt}}
\put(298.0,131.0){\rule[-0.200pt]{0.400pt}{4.818pt}}
\put(298,90){\makebox(0,0){-8000}}
\put(298.0,756.0){\rule[-0.200pt]{0.400pt}{4.818pt}}
\put(425.0,131.0){\rule[-0.200pt]{0.400pt}{4.818pt}}
\put(425,90){\makebox(0,0){-7000}}
\put(425.0,756.0){\rule[-0.200pt]{0.400pt}{4.818pt}}
\put(551.0,131.0){\rule[-0.200pt]{0.400pt}{4.818pt}}
\put(551,90){\makebox(0,0){-6000}}
\put(551.0,756.0){\rule[-0.200pt]{0.400pt}{4.818pt}}
\put(678.0,131.0){\rule[-0.200pt]{0.400pt}{4.818pt}}
\put(678,90){\makebox(0,0){-5000}}
\put(678.0,756.0){\rule[-0.200pt]{0.400pt}{4.818pt}}
\put(805.0,131.0){\rule[-0.200pt]{0.400pt}{4.818pt}}
\put(805,90){\makebox(0,0){-4000}}
\put(805.0,756.0){\rule[-0.200pt]{0.400pt}{4.818pt}}
\put(932.0,131.0){\rule[-0.200pt]{0.400pt}{4.818pt}}
\put(932,90){\makebox(0,0){-3000}}
\put(932.0,756.0){\rule[-0.200pt]{0.400pt}{4.818pt}}
\put(1059.0,131.0){\rule[-0.200pt]{0.400pt}{4.818pt}}
\put(1059,90){\makebox(0,0){-2000}}
\put(1059.0,756.0){\rule[-0.200pt]{0.400pt}{4.818pt}}
\put(1185.0,131.0){\rule[-0.200pt]{0.400pt}{4.818pt}}
\put(1185,90){\makebox(0,0){-1000}}
\put(1185.0,756.0){\rule[-0.200pt]{0.400pt}{4.818pt}}
\put(1312.0,131.0){\rule[-0.200pt]{0.400pt}{4.818pt}}
\put(1312,90){\makebox(0,0){ 0}}
\put(1312.0,756.0){\rule[-0.200pt]{0.400pt}{4.818pt}}
\put(1439.0,131.0){\rule[-0.200pt]{0.400pt}{4.818pt}}
\put(1439,90){\makebox(0,0){ 1000}}
\put(1439.0,756.0){\rule[-0.200pt]{0.400pt}{4.818pt}}
\put(171.0,131.0){\rule[-0.200pt]{0.400pt}{155.380pt}}
\put(171.0,131.0){\rule[-0.200pt]{305.461pt}{0.400pt}}
\put(1439.0,131.0){\rule[-0.200pt]{0.400pt}{155.380pt}}
\put(171.0,776.0){\rule[-0.200pt]{305.461pt}{0.400pt}}
\put(30,453){\makebox(0,0){\rotatebox{90}{Counts}}}
\put(805,29){\makebox(0,0){Value}}
\put(805,838){\makebox(0,0){$t_{\bar{\cal C}}$}}
\put(250.0,131.0){\rule[-0.200pt]{0.400pt}{0.482pt}}
\put(250.0,133.0){\rule[-0.200pt]{3.854pt}{0.400pt}}
\put(266.0,131.0){\rule[-0.200pt]{0.400pt}{0.482pt}}
\put(250.0,131.0){\rule[-0.200pt]{3.854pt}{0.400pt}}
\put(871.0,131.0){\rule[-0.200pt]{0.400pt}{0.482pt}}
\put(871.0,133.0){\rule[-0.200pt]{4.095pt}{0.400pt}}
\put(888.0,131.0){\rule[-0.200pt]{0.400pt}{0.482pt}}
\put(871.0,131.0){\rule[-0.200pt]{4.095pt}{0.400pt}}
\put(1002.0,131.0){\rule[-0.200pt]{0.400pt}{0.482pt}}
\put(1002.0,133.0){\rule[-0.200pt]{4.095pt}{0.400pt}}
\put(1019.0,131.0){\rule[-0.200pt]{0.400pt}{0.482pt}}
\put(1002.0,131.0){\rule[-0.200pt]{4.095pt}{0.400pt}}
\put(1035.0,131.0){\rule[-0.200pt]{0.400pt}{0.482pt}}
\put(1035.0,133.0){\rule[-0.200pt]{3.854pt}{0.400pt}}
\put(1051.0,131.0){\rule[-0.200pt]{0.400pt}{0.482pt}}
\put(1035.0,131.0){\rule[-0.200pt]{3.854pt}{0.400pt}}
\put(1051.0,131.0){\rule[-0.200pt]{0.400pt}{0.482pt}}
\put(1051.0,133.0){\rule[-0.200pt]{4.095pt}{0.400pt}}
\put(1068.0,131.0){\rule[-0.200pt]{0.400pt}{0.482pt}}
\put(1051.0,131.0){\rule[-0.200pt]{4.095pt}{0.400pt}}
\put(1068.0,131.0){\rule[-0.200pt]{0.400pt}{1.445pt}}
\put(1068.0,137.0){\rule[-0.200pt]{3.854pt}{0.400pt}}
\put(1084.0,131.0){\rule[-0.200pt]{0.400pt}{1.445pt}}
\put(1068.0,131.0){\rule[-0.200pt]{3.854pt}{0.400pt}}
\put(1084.0,131.0){\rule[-0.200pt]{0.400pt}{0.482pt}}
\put(1084.0,133.0){\rule[-0.200pt]{3.854pt}{0.400pt}}
\put(1100.0,131.0){\rule[-0.200pt]{0.400pt}{0.482pt}}
\put(1084.0,131.0){\rule[-0.200pt]{3.854pt}{0.400pt}}
\put(1100.0,131.0){\rule[-0.200pt]{0.400pt}{1.445pt}}
\put(1100.0,137.0){\rule[-0.200pt]{4.095pt}{0.400pt}}
\put(1117.0,131.0){\rule[-0.200pt]{0.400pt}{1.445pt}}
\put(1100.0,131.0){\rule[-0.200pt]{4.095pt}{0.400pt}}
\put(1117.0,131.0){\rule[-0.200pt]{0.400pt}{2.650pt}}
\put(1117.0,142.0){\rule[-0.200pt]{3.854pt}{0.400pt}}
\put(1133.0,131.0){\rule[-0.200pt]{0.400pt}{2.650pt}}
\put(1117.0,131.0){\rule[-0.200pt]{3.854pt}{0.400pt}}
\put(1133.0,131.0){\rule[-0.200pt]{0.400pt}{0.964pt}}
\put(1133.0,135.0){\rule[-0.200pt]{3.854pt}{0.400pt}}
\put(1149.0,131.0){\rule[-0.200pt]{0.400pt}{0.964pt}}
\put(1133.0,131.0){\rule[-0.200pt]{3.854pt}{0.400pt}}
\put(1149.0,131.0){\rule[-0.200pt]{0.400pt}{4.095pt}}
\put(1149.0,148.0){\rule[-0.200pt]{4.095pt}{0.400pt}}
\put(1166.0,131.0){\rule[-0.200pt]{0.400pt}{4.095pt}}
\put(1149.0,131.0){\rule[-0.200pt]{4.095pt}{0.400pt}}
\put(1166.0,131.0){\rule[-0.200pt]{0.400pt}{6.263pt}}
\put(1166.0,157.0){\rule[-0.200pt]{3.854pt}{0.400pt}}
\put(1182.0,131.0){\rule[-0.200pt]{0.400pt}{6.263pt}}
\put(1166.0,131.0){\rule[-0.200pt]{3.854pt}{0.400pt}}
\put(1182.0,131.0){\rule[-0.200pt]{0.400pt}{7.950pt}}
\put(1182.0,164.0){\rule[-0.200pt]{4.095pt}{0.400pt}}
\put(1199.0,131.0){\rule[-0.200pt]{0.400pt}{7.950pt}}
\put(1182.0,131.0){\rule[-0.200pt]{4.095pt}{0.400pt}}
\put(1199.0,131.0){\rule[-0.200pt]{0.400pt}{8.913pt}}
\put(1199.0,168.0){\rule[-0.200pt]{3.854pt}{0.400pt}}
\put(1215.0,131.0){\rule[-0.200pt]{0.400pt}{8.913pt}}
\put(1199.0,131.0){\rule[-0.200pt]{3.854pt}{0.400pt}}
\put(1215.0,131.0){\rule[-0.200pt]{0.400pt}{15.177pt}}
\put(1215.0,194.0){\rule[-0.200pt]{3.854pt}{0.400pt}}
\put(1231.0,131.0){\rule[-0.200pt]{0.400pt}{15.177pt}}
\put(1215.0,131.0){\rule[-0.200pt]{3.854pt}{0.400pt}}
\put(1231.0,131.0){\rule[-0.200pt]{0.400pt}{28.908pt}}
\put(1231.0,251.0){\rule[-0.200pt]{4.095pt}{0.400pt}}
\put(1248.0,131.0){\rule[-0.200pt]{0.400pt}{28.908pt}}
\put(1231.0,131.0){\rule[-0.200pt]{4.095pt}{0.400pt}}
\put(1248.0,131.0){\rule[-0.200pt]{0.400pt}{44.326pt}}
\put(1248.0,315.0){\rule[-0.200pt]{3.854pt}{0.400pt}}
\put(1264.0,131.0){\rule[-0.200pt]{0.400pt}{44.326pt}}
\put(1248.0,131.0){\rule[-0.200pt]{3.854pt}{0.400pt}}
\put(1264.0,131.0){\rule[-0.200pt]{0.400pt}{101.178pt}}
\put(1264.0,551.0){\rule[-0.200pt]{3.854pt}{0.400pt}}
\put(1280.0,131.0){\rule[-0.200pt]{0.400pt}{101.178pt}}
\put(1264.0,131.0){\rule[-0.200pt]{3.854pt}{0.400pt}}
\put(1280.0,131.0){\rule[-0.200pt]{0.400pt}{149.599pt}}
\put(1280.0,752.0){\rule[-0.200pt]{4.095pt}{0.400pt}}
\put(1297.0,131.0){\rule[-0.200pt]{0.400pt}{149.599pt}}
\put(1280.0,131.0){\rule[-0.200pt]{4.095pt}{0.400pt}}
\put(1297.0,131.0){\rule[-0.200pt]{0.400pt}{101.178pt}}
\put(1297.0,551.0){\rule[-0.200pt]{3.854pt}{0.400pt}}
\put(1313.0,131.0){\rule[-0.200pt]{0.400pt}{101.178pt}}
\put(1297.0,131.0){\rule[-0.200pt]{3.854pt}{0.400pt}}
\put(171.0,131.0){\rule[-0.200pt]{0.400pt}{155.380pt}}
\put(171.0,131.0){\rule[-0.200pt]{305.461pt}{0.400pt}}
\put(1439.0,131.0){\rule[-0.200pt]{0.400pt}{155.380pt}}
\put(171.0,776.0){\rule[-0.200pt]{305.461pt}{0.400pt}}
\end{picture}} 
\et
\ec
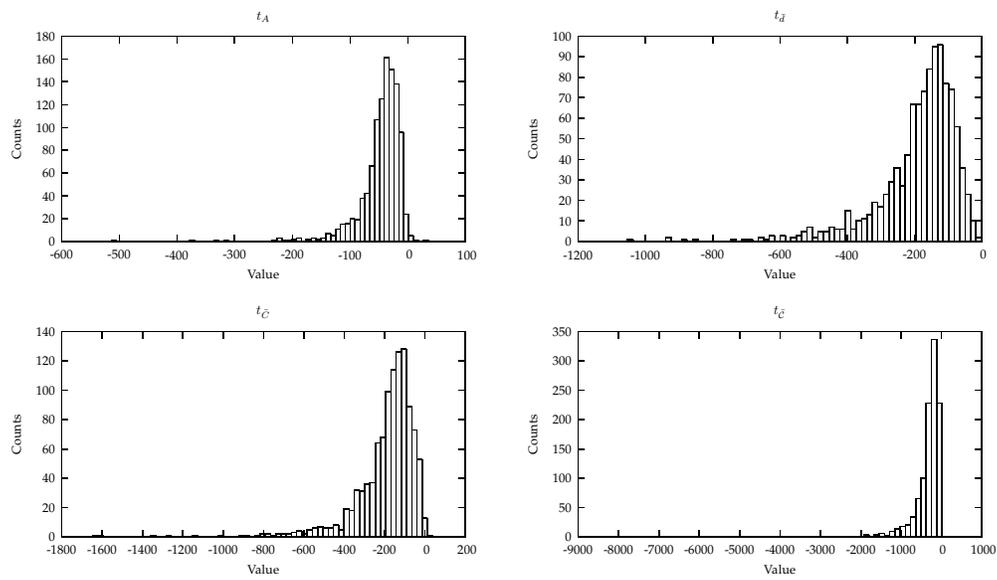
\caption{Histograms for parameter deviations in PPI networks. Top left, right: assortativity and average shortest path. Bottom left, right: Average local clustering and global clustering.} \label{fig2rand}
\efigs

In the case of PPI networks, the differences between real and expected values for all parameters are statistically significant. Real PPI networks present bigger average local and global clustering coefficients than their expected values in random networks with the same degree distribution, as well as bigger shortest average paths and assortativity. Realistic evolutionary models that try to mimic growth of PPI networks should not only try to reproduce their degree distribution (which is not well described by a power-law function) but also try to incorporate underling mechanisms that result in networks with such deviations from the expected values in networks with the same degree distributions.


\section{Overview and Conclusion}

We have analyzed a huge set of graphs representing biological systems (3481 metabolic networks and 1073 PPI networks). First, we study in detail the degree distributions of the networks and test them against the hypothesis that they follow a power-law function. The results of this first analysis show that the degree distributions of these real world graphs are not well described by this function, but only the right tail containing a smaller fraction of the nodes in the case of PPI networks are reasonably adjusted to this scale-free distribution. In a second analysis, a complete descriptive statistics of the networks properties is presented and then we identify those parameters that deviate from their expected values in randomized versions of the graphs that preserve the network's degree distribution. Biological networks are the result of evolution and natural selection. Therefore, these deviations are the result of evolutionary pressures these systems developed under. Realistic evolutionary models that describe these systems should incorporate mechanisms that result in graphs with such deviations and not only try to reproduce their degree distributions. 

Our analysis did not focus in any given specific branch of the tree of life, such that the networks vary a lot in size. The average shortest path for both, metabolic and PPI networks, tend to be slightly smaller in the case of real graphs than the expected values in randomized networks with high statistical significance. For other parameters, while the metabolic networks also show some differences, these differences do not present the same confidence level significance over the whole sample. But in the case of PPI networks, all parameters analyzed do show differences between real and expected values with a high confidence level. PPI networks are more assortative, and have bigger local and global clustering coefficients than would be expected by randomly linking nodes with the same degree distributions.

This study points to two important conclusions: on one hand, the degree distributions of graphs representing biological systems (metabolic and PPI networks) is not well fitted by a power-law function and on the other hand, network evolutionary models that focus solely in obtaining graphs with similar degree distributions as real world biological networks might not be enough to adjust other graph parameters observed in these systems.


\begin{appendices}

\section{Parameter Distribution over the Randomized Networks}

In this work we compared the graph parameters of a real world network with their expected values in the set of all graphs that can be built with the same degree distribution. This analysis is inspired by the fact that many models that try to describe the evolution of these systems usually focus in obtaining as a result of the model simulation graphs with similar degree distributions. We ask, therefore, if mimicking this degree distribution is enough to reproduce the main properties of these structures. Given a degree distribution, the number of different graphs that can be actually built from it is astronomically huge and a random model adjusted only to reproduce a given degree distribution would, in principle, generate any of the possible networks that share this same distribution. So the question is: Do the characteristics of a real world network significantly differ from those of a randomly selected graph from the population of all networks that share the same degree distribution?

The process through which we obtain samples from this population is the randomization process described in section \ref{randana}, where links are broken and rewired in order to obtain a complete new network from the original but keeping the degrees of the nodes untouched. Though this rewiring process is computationally fast, the computation of all parameters of the resulting network, in particular $P_2$, $C_3$ and $d_{ij}$, is a computationally intense process and in order to perform all calculations in a relatively short and reasonable time period (around a month) we chose to obtain a small random sample. Therefore, we had to use a statistical test designed to provide reliable results even for small samples: the student's t-test. One of the suppositions behind the student's t-test performed in the analysis is that the population behind the obtained sample has a normal distribution.

Here, for a few organisms, we obtained bigger random samples and performed two normality tests to check whether these populations comply with the needed supposition of normality in order to perform the t-test. In table \ref{tab:normtest} we present the results for the PPI and metabolic networks of a few selected organisms. The table shows the average value of the given parameters in a sample of randomized networks and for each parameter two p-values next to it that are the result of the Shapiro-Wilk \cite{shapiro} and D'Agostino \cite{dagostino} normality tests. In any of the tests, a big p-value (bigger than the critical value) indicates that the sample seems to have a normal distribution. In the tables, for the PPI networks we identify the organisms by their NCBI taxonomy ID and metablic networks by their KEGG code. Again, in this evaluation we reduced the sample sizes for those networks in which the calculations took longer time. 

\btabs[h]
\caption{Normality test for a few selected organisms. The average value of the parameters over the samples are shown and the two values next to each are the p-value of the Shapiro-Wilk test (pv1) and the p-value for the D'agostino test (pv2). } \label{tab:normtest}
\bc
\bt{cccc|cccc}
Organism & $N$ & $\cal N$ & Sample Size & $\bar{C}$ (pv1, pv2) & $A$ (pv1, pv2) & $\bar{d}$ (pv1, pv2) & $\cal{C}$ (pv1, pv2) \\
\hline
mgl &   823 &  2416 &   500 & 0.190693 (0.458, 0.510) & -0.230871 (0.016, 0.018) & 3.020923 (0.483, 0.799) & 0.054189 (0.170, 0.661) \\
syn &   967 &  2651 &   500 & 0.187642 (0.108, 0.138) & -0.211546 (0.350, 0.406) & 3.058312 (0.510, 0.365) & 0.041776 (0.482, 0.715) \\
eco &  1195 &  3564 &   500 & 0.192970 (0.399, 0.403) & -0.204386 (0.406, 0.227) & 3.010009 (0.245, 0.145) & 0.037497 (0.405, 0.275) \\
243273 &   436 &  6107 &   500 & 0.137612 (0.290, 0.216) & -0.046098 (0.134, 0.165) & 2.200834 (0.124, 0.069) & 0.132851 (0.866, 0.860) \\
511145 &  4076 & 70424 &   100 & 0.041745 (0.208, 0.256) & -0.044631 (0.271, 0.586) & 2.639252 (0.315, 0.697) & 0.033799 (0.470, 0.410) \\
4932 &  6018 & 252957 &   100 & 0.046844 (0.428, 0.597) & -0.037580 (0.075, 0.181) & 2.300632 (0.717, 0.538) & 0.041193 (0.449, 0.342) 
\et    
\ec
\etabs

\end{appendices}


\bibliographystyle{ieeetr}
\bibliography{biographs}

\end{document}